\documentclass[pre,twocolumn,superscriptaddress,showkeys]{revtex4-2}
\usepackage[english]{babel}
\usepackage[utf8]{inputenc}
\usepackage{graphicx}
\usepackage{ragged2e}
\usepackage{amsmath}
\usepackage[hidelinks]{hyperref}

\begin{document}
\title{Paradoxes in the co-evolution of contagions and institutions}

\author{
\firstname{Jonathan}
\surname{St-Onge}
}
\email{These two authors contributed equally}

\affiliation{
  Vermont Complex Systems Center, University of Vermont, Burlington VT, USA.
}

\author{
\firstname{Giulio}
\surname{Burgio}
}
\email{These two authors contributed equally}

\affiliation{
  Departament d'Enginyeria Inform\`atica i Matem\`atiques, Universitat Rovira i Virgili, 43007 Tarragona, Spain.
}

\author{
\firstname{Samuel}
\surname{F. Rosenblatt}
}

\affiliation{
  Vermont Complex Systems Center, University of Vermont, Burlington VT, USA.
}

\affiliation{
  Department of Computer Science, University of Vermont, Burlington VT, USA.
}

\author{
\firstname{Timothy}
\surname{M. Waring}
}

\affiliation{
  School of Economics, University of Maine, Orono ME, USA.
}

\affiliation{
  Mitchell Center for Sustainability Solutions, University of Maine, Orono ME, USA.
}

\author{
\firstname{Laurent}
\surname{H\'ebert-Dufresne}
}
\email{laurent.hebert-dufresne@uvm.edu}

\affiliation{
  Vermont Complex Systems Center, University of Vermont, Burlington VT, USA.
}

\affiliation{
  Department of Computer Science, University of Vermont, Burlington VT, USA.
}

\date{\today} 

\begin{abstract}
Epidemic models study the spread of an undesired agent through a population, be it infectious diseases through a country, misinformation in online social media, or pests infesting a region. 
In combating these epidemics, we rely neither on global top-down interventions, nor solely on individual adaptations. 
Instead, interventions most commonly come from local institutions such as public health departments, moderation teams on social media platforms, or other forms of group governance. 
Classic models, which are often individual or agent-based, are ill-suited to capture local adaptations.
We leverage recent development of institutional dynamics based on cultural group selection to study how groups can attempt local control of an epidemic by taking inspiration from the successes and failures of other groups. 
Incorporating these institutional changes into the epidemic dynamics reveals paradoxes: a higher transmission rate can result in smaller outbreaks and decreasing the speed of institutional adaptation generally reduces outbreak size. 
When groups perceive a contagion as more worrisome, they can invest in improved policies and, if they maintain these policies long enough to have impact, lead to a reduction in endemicity. 
By looking at the interplay between the speed of institutions and the transmission rate of the contagions,
we find rich co-evolutionary dynamics that reflect the complexity of known biological and social contagions.
\end{abstract}

\keywords{Group-based master equations, coevolution, institutions, collective decision-making, cultural evolution, contagions}

\maketitle

\section{Introduction}
\label{sec:papertag.introduction}

Epidemic models are increasingly integrating social and behavioural factors to elucidate the complex dynamics of various types of contagions. However, in most cases these coupled dynamics track the individual level at which an agent spreads through host units (e.g., people, users, or locations), disregarding the higher collective level at which interventions are decided and implemented. 
Consider different policies used to limit the spread of SARS-CoV-2, such as mask mandates and gathering closures, which were not enforced at a global level but often implemented at the level of towns, counties, or states, in very inconsistent ways \cite{althouse_unintended_2020}. As a second example, consider how social contagions like misinformation or hate speech unfold heterogeneously on social media across communities which set their own norms and moderation policies (e.g., subreddits on Reddit, or instances on Mastodon). Existing models seldom consider the mesoscopic scale at which these policies are set, and even more rarely consider how these policies might co-evolve with the contagion dynamics as collectives attempt to copy each other based on perceived success.

By tracking groups instead of individuals, we can investigate how changes in group norms or policies can hinder contagion. Those changes create a negative feedback loop between the contagions and institutions. Groups impose policies about the right behaviours that group members should adopt during a contagion, without intervening at the level of network structure. More effective policies can be adopted, albeit at a greater cost (e.g., coordination costs, loss of members to other groups,  paying a political price). 

In the case of SARS-CoV-2, we hypothesise that the array of institutional responses emerged due to interaction of group dynamics, where institutions emulated each other's policies in their attempts to address contagion. We assume that groups gauge their policy strength by evaluating their own performance in managing local infection rates in comparison to groups with stronger or weaker policies. For instance, a public health department might enforce strong preventive behaviours, reducing the global prevalence rate. Under a certain level of intergroup coupling, a weaker public health department might forego the imposition of stronger policies if it views the situation under control, thereby saving the cost of enforcement. This situation exemplifies \textit{institutional free riding}, in which weaker institutions benefit from the investment made by stronger ones.

Group-based master equations (GMEs \cite{hebert-dufresne_propagation_2010,st-onge_master_2021,hebert-dufresne_source-sink_2022}) allow us to examine the unfolding of policies on networks with higher-order organisation. GMEs follow the probability for groups to find themselves in any of the available discrete states; with regards to both contagion (e.g. prevalence) and institutions (see Fig. \ref{fig:cartoon_model}). Effective policies in this context translate into a lower probability of moving to a state with a larger number of infected members (as policies improve, you would tend to observe less infected members in Fig. \ref{fig:cartoon_model}). GMEs differ both from mean-field methods that follow the macroscopic, average behaviour of the system, and the complexities of following individual agents.

The paper proceeds as follows. We review a curated list of related works encompassing co-evolutionary dynamics of contagion and adaptation. We show how our group-based approach is related but distinct from other epidemiological work based on on co-evolutionary dynamics, public good games in environmental studies, and the innovation patterns as studied by sociologists and economists. We introduce our model and results in the subsequent sections. Throughout the paper, we make the methodological argument that adopting a group-based approach to model the co-evolution of contagions and institutions actually makes the problem more tractable than sticking to an individual perspective. 

\section{Related works}
\label{sec:papertag.related}

Before undertaking this brief review, we set forth three features that encompass the co-evolutionary processes expressed by our formalism. 
Firstly, we model a multi-scale process coupling two dynamics; one at the level of individuals and one at the collective level of groups.
Secondly, groups can act more effectively on the individual dynamics by embracing institutions/policies of increasing strength, albeit at a greater collective cost.
Thirdly, groups copy each other based on their perceived fitness. The latter depends on both the state of the members of a group in relation to a spreading entity/behaviour and the cost entailed by the institutions/policies adopted by the group. Hence, there is a group-level traits (institutions/policies) selection revolving around the collective dilemma of maximizing joint welfare while minimizing collective costs.
We examine works from various fields that relate to these three features. 

\paragraph*{Contagions as common-pool resources.} Public good games (PGGs) in environmental studies is a key area of research studying social dilemmas based on joint welfare. It represents a situation where if individuals were willing to personally bear a cost for the benefit of the collective good, it would be advantageous for everyone involved. But humans thought as rational actors fail to do so. In environmental studies, this is conventionally framed as the ``tragedy of the commons''~\cite{olson_logic_1971, ostrom_governing_1990}.

By seeing contagion as a kind of environmental good that must not be harvested to be beneficial to the groups, we can understand the co-evolution of contagions and institutions as a collective action problem. In this context, we can frame behavioural strategies to be a social dilemma between \textsc{rest} (avoiding the risk to harvest the good, hence cooperating) and \textsc{go out} (exposing themselves to that risk, hence defecting). 

Institutional theorists rebutted  the tragedy of the commons by showcasing groups' abilities to adopt rules/policies to govern the commons, both from grassroots initiatives and top-down approaches~\cite{ostrom_governing_1990}. They made extensive use of game theory to empirically study the success and failures to manage public goods under different sets of rules. They found that the degree of success of both public and private institutions varies widely across contexts, and that getting the institutions right is the real challenge. PGGs can depict the diversity of interactions between individuals and institutions in policy-making, but they have not yet captured the evolving dynamics as policy strength varies over time.

\paragraph*{Contagions as innovations.} 

The study of innovation is another key research area that explores the multi-scale process of how firms, innovators, and markets interact to produce and diffuse innovations \cite{schumpeter_theory_1934, metcalfe_evolutionary_1998, fagerberg_innovation_2006}. Our model aligns most closely with the literature on diffusion of innovation. For instance, acquiring electric vehicles (EVs) entails greater financial costs when compared to fossil fuel-powered cars. However, a broader adoption of EVs would have the collective benefit of reducing greenhouse gas emissions. When discussing consumers being incentivised by government to purchase EVs, the study of institutions promoting diffusion of costly innovations can be seen as the inverse of institutions seeking to prevent the spread of contagions.

Our formalism encompasses the study of innovation patterns as well, but only when the coevolutionary process adheres to the following; whether innovators are the ones who benefit from the adoption of the innovations, if there is a cost incurred with adopting the innovation, and whether institutions promoting the innovation represent the adopters. Consider the development of Free and Open-Source Software (F/OSS). In this case, the innovators themselves are also the adopters, the adoption of F/OSS can be considered costly compared to using proprietary software from corporations, and institutions have emerged to promote the use of F/OSS among groups. When innovators are distinct from the consumers, and firms are third-party actors who seek to mediate how innovation spreads in the markets, this falls outside the scope of our formalism

An analogy can be drawn between the key concept of creative destruction in the realm of long-term economic change and the trade-offs linked to fast and slow institutional copying rates in times of contagion. Creative destruction is the notion that older, well-established firms become entrenched in routine technological systems (deepening) and face the risk of being supplanted by smaller, more adaptable firms that can better respond to external changes~\cite{ breschi_technological_2000, fagerberg_innovation_2006}. There is a trade-off between reactivity to the outside knowledge and ''organisational memory'' embedded in how firms can make their operations run smoothly~\cite{cohen_chapter_1989}. That said, this analogy is imperfect as we don't have heterogeneity of firm size in our model. 

Innovation studies have explored the diffusion of innovations at the microscopic level, the role of innovators in macroscopic technological competitions, and the significance of the coordination between firms and consumers in economic performance. 
A key difference in our model from the previous literature is that we only model the co-evolution between institutions that can intervene locally on individual behaviours, and ignore third-party actors (e.g. corporations, lobbyists) that can mediate the spreading entity.

\paragraph*{Coupled contagion models.} One last related perspective from epidemiology is to model how individual beliefs are entangled with that of contagion dynamics. Works on duelling~\cite{fu_dueling_2017}, parasitic~\cite{hebert-dufresne_spread_2020} or antagonistic~\cite{smaldino_coupled_2021} contagion all share the idea that there is an informational layer---what individuals believe about the contagion, or the underlying social identities influencing behaviours---that facilitate or not the contagion. One way to formally adopt this socio-ecological perspective is to use two-layered networks to encode the structure of the two kinds of interactions at play.

Importantly, in most coupled contagion models, intervention is an individual and contagious process, not a group process~\cite{WANG20151, bedson2021review}. This does not imply that groups are insignificant in the context of contagion within this framework. For instance, Smaldino \& Jones assume that their informational layer depends on outgroup aversion \cite{smaldino_coupled_2021}. This, in turn, is assumed to slow down the rate at which, say, one group adopts masking behaviours if the behaviour is associated with an adverse group. 

While individual behaviours may be influenced by group affiliations, focusing on individuals rather than the groups themselves tends to neglect the role of institutions in shaping the norms that guide individual behaviours. At the risk of stating the obvious, modelling groups instead of individuals here facilitate the study of this feedback loop given policies are implemented at the group level.

Our review of prior work illustrates two key points. First, we found important insights about the role and dynamics of institutions in classical studies on PGGs and patterns of innovation from environmental studies and economics, respectively. In particular, those fields highlight the importance of understanding how groups mediate the diffusion of behavioral strategies or innovations at the individual level. Second, despite clear parallels, institutional dynamics is surprisingly absent from current research on interventions against harmful contagions such as epidemics, misinformation, or pests. We aim to fill this gap by incorporating institutional dynamics in a group-based epidemic model. 

\section{Model}
\label{sec:papertag.model}

We follow a previous model for the co-evolution of collective behaviour (e.g., cooperation) and institutions~\cite{hebert-dufresne_source-sink_2022}, but here assume that institutions are meant to limit the spread of epidemics rather than promote positive behaviours. We consider a large population of $N \gg 1$ host units, structured as $M$ groups of size $n = N/M$. Host units could be, for example, individuals who can get infected with a pathogen, digital accounts that can be compromised, or locations that can be infested. We consider a simple susceptible-infectious-susceptible (SIS) dynamic where susceptible host units get the infection at a transmission rate $\beta_0$ from any infectious agent, either from within their group or from other groups, and infected units becomes susceptible again at a recovery rate $\gamma$.

\begin{figure*}[t]
  \centering	
    \includegraphics[width=0.75\linewidth]{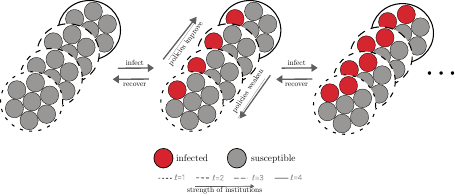}  
  \caption{  \justifying 
    \textbf{Graphical representation of the model.}
    Epidemic (left-right) and selection (forward-backward) transitions between the possible states of a group as tracked by our set of group-based master equations, Eqs.~(\ref{eq:total_me}). Groups all have the same size and are distinguished by the number of infected members $i$ and by the institutional level $\ell$.
  }
  \label{fig:cartoon_model}
\end{figure*}

Groups can put in place norms and policies in order to limit contagion. Such interventions can be of various kinds and intensities and, accordingly, have different efficacy in avoiding infections. We capture such group-level institutional variability through an abstract institutional level $\ell\in\{1,\dots, L\}$ which translates to a local transmission rate $\beta_\ell = \beta_0\ell^{-\alpha}$, with $\alpha>0$, such that a larger $\ell$ implies a smaller rate. Accordingly, $\beta_1=\beta_0$, meaning that no intervention is implemented at level $\ell = 1$. The parameter $\alpha$ quantifies how institutional investment impacts on transmission (with diminishing returns, i.e., the decrease in $\beta_\ell$ becomes increasingly smaller by raising $\ell$). Whether through contacts with individuals of their own group or from another group, members of a group of level $\ell$ are infected at a rate $\beta_\ell$. The institutional level of a group is thus equivalent to both (1) a reduction in local transmission (e.g., ventilation against airborne pathogens, or moderation of online content) or (2) a modification of the susceptibility of individuals (e.g., prophylaxis against a disease, or education against misinformation). The model is thus abstract enough to encompass a variety of real situations, providing in this way a general framework to study the intertwined dynamics of institutions and individuals.

We employ a set of group-based master equations to model the epidemic dynamics across groups \cite{hebert-dufresne_propagation_2010}. GMEs track the fraction $G_{i,\ell}(t)$ of groups with $i$ infectious units and institutional level $\ell$ at time $t$. Capturing the distribution of groups across both dimensions allows us to capture their recent history such that, for example, groups that increase their institutional level do not immediately see a decrease in infection. Over this two-dimensional space, the epidemic dynamics follow
\begin{equation} \label{eq:diffusion_me}
    \begin{split}
        \frac{d}{dt}G_{i,\ell}^{\textrm{epi}} &= \beta_\ell\left[(i-1)+R\right]\left(n-i+1\right)G_{i-1,\ell} \\
        &- \beta_\ell\left(i+R\right)\left(n-i\right)G_{i,\ell} \\
        &+ \gamma\left(i+1\right)G_{i+1,\ell} - \gamma i G_{i,\ell} \; . 
    \end{split}
\end{equation}
This master equation tracks the evolution of the distribution of all possible group states by accounting for the transitions between any two adjacent states. Eq.~(\ref{eq:diffusion_me}) tracks the epidemic dynamics only (hence the `epi' superscript). The first term corresponds to a transmission event bringing groups from state $(i-1,\ell)$ to state $(i,\ell)$, which occurs proportionally to the internal transmission rate $\beta_\ell = \beta_0\ell^{-\alpha}$ multiplied by the number of susceptible units $n-(i-1)$ that could be infected. The factor in square brackets corresponds to the number of contacts these susceptible units have either with the $i-1$ infectious units in their group and possibly with infectious units in other groups, which we set equal to

\begin{equation}
R = \rho \sum _{i,\ell} i G_{i,\ell} \; .
\end{equation}

This is written assuming random mixing between groups, therefore as the product of an inter-group coupling $\rho$ and the expected number of infected units in a randomly selected group ($\sum _{i,\ell} i G_{i,\ell}$).
The second term in Eq.~(\ref{eq:diffusion_me}) also corresponds to transmission events, but now taking groups out of state $(i,\ell)$ into state $(i+1,\ell)$. Finally, the last two terms correspond to recovery events; bringing groups from state $(i+1,\ell)$ to $(i,\ell)$ and from $(i,\ell)$ to $(i-1,\ell)$.

Additionally, we model institutional dynamics among groups. A higher institutional strength better reduces transmission, but also implies greater institutional effort and resources. Groups thus test whether an institutional investment to prevent infections is worth its cost. Assigned a perceived fitness to groups, we assume that a group pays a fixed cost $c >0 $ for each additional level of institutional strength, but it is penalised by a fixed amount $b > 0$ for each infected unit within it. Institutions then evolve via a selection process in which groups choose their institutional level based on the average perceived fitness associated to that level, here defined as
\begin{equation}
Z_\ell = \dfrac{\sum_{i} \textrm{exp}\left(-bi - c\ell\right) G_{i,\ell}}{\sum_{i} G_{i,\ell}} \; .
\end{equation}
The selection process at the group-level is thus described by a second master equation,
\begin{align}
    \label{eq:selection_me}
        \frac{d}{dt}G_{i,\ell}^{\textrm{sel}} &= \eta \left[G_{i,\ell-1}\left(Z_\ell Z_{\ell-1}^{-1} + \mu\right) + G_{i,\ell+1}\left(Z_\ell Z_{\ell+1}^{-1}+\mu\right) \right] \notag \\
         &- \eta \left(Z_{\ell - 1}Z_{\ell}^{-1}+Z_{\ell+1}Z_{\ell}^{-1} + 2\mu\right)G_{i,\ell}\; . 
\end{align}
tracking flows of groups from institutional strength $\ell$ to $\ell+1$ and $\ell-1$ (and vice versa). These flows are assumed to occur proportionally to the relative fitness of different levels (e.g., a group moves from $\ell$ to $\ell +1 $ proportionally to $Z_{\ell+1} / Z_{\ell}$) times the rate of institutional copying rate $\eta$. We also add a fitness-independent rate of spontaneous institutional change $\mu$ that allows to vary the selection strength, which is maximal for $\mu = 0$ and zero for $\mu\rightarrow\infty$ (with $\eta\rightarrow 0$ such that $\eta\mu$ is finite).

Equation~(\ref{eq:selection_me}) is certainly only one of several possible models approximating how institutions might co-evolve with the contagion dynamics. It represents naive institutions that follow a random walk in institutional space which is biased by levels' fitness, yet disregarding other factors such as the history of the groups they copy or whether interventions require resources to be implemented. Importantly, our model allows to easily incorporate different ways in which institutions react to changing environments. When this happen, there is a dance of who decides to copy whom, influencing the relative benefits of group-level social learning. As with individual-level learning strategies, we could supplement our model with other social learning strategies, where groups might imitate other groups based on cultural similarity or simply conform to the majority \cite{boyd_culture_1985}. 

Regardless of specific modelling choices, the overall dynamics of our model is tracked by combining the epidemic and selection dynamics in the following set of master equations,

\begin{equation}
\label{eq:total_me}
    \frac{d}{dt}G_{i,\ell}= \frac{d}{dt}G_{i,\ell}^{\textrm{epi}} + \frac{d}{dt}G_{i,\ell}^{\textrm{sel}} \; ,
\end{equation}
defined over $i\in\{0,\dots, n\}$ and $\ell\in\{1,\dots, L\}$ (see Fig.~\ref{fig:cartoon_model} for a graphical representation). We can thus follow the dynamics of our system by numerically integrating the equations starting from arbitrary initial conditions (constrained only by $\sum_{i,\ell} G_{i,\ell} = 1$).

\section{Results}
\label{sec:papertag.results}

Using GMEs allows us to obtain a detailed description of the system by tracking the evolution of the entire distribution of group states, while evading the computational cost entailed by numerous agent-based stochastic simulations.
The model takes as input a set of values for the parameters introduced in the previous section ($n$, $\beta_0$, $\gamma$, $\rho$, $\eta$, and $\mu$), and then produces non-trivial dynamics as we numerically track Eq.~(\ref{eq:total_me}). The output can be analysed either as (sec. \ref{sec:papertag.tempAnalysis}) time series of global prevalence across the entire population or within institutions of a specific level along with time series of the popularity of specific institutional levels, or (sec. \ref{sec:eq}) by looking at the dynamical equilibrium (or fixed point) the system eventually reaches.

\begin{figure*}[tp!]
    \centering
    \includegraphics[width=\linewidth]{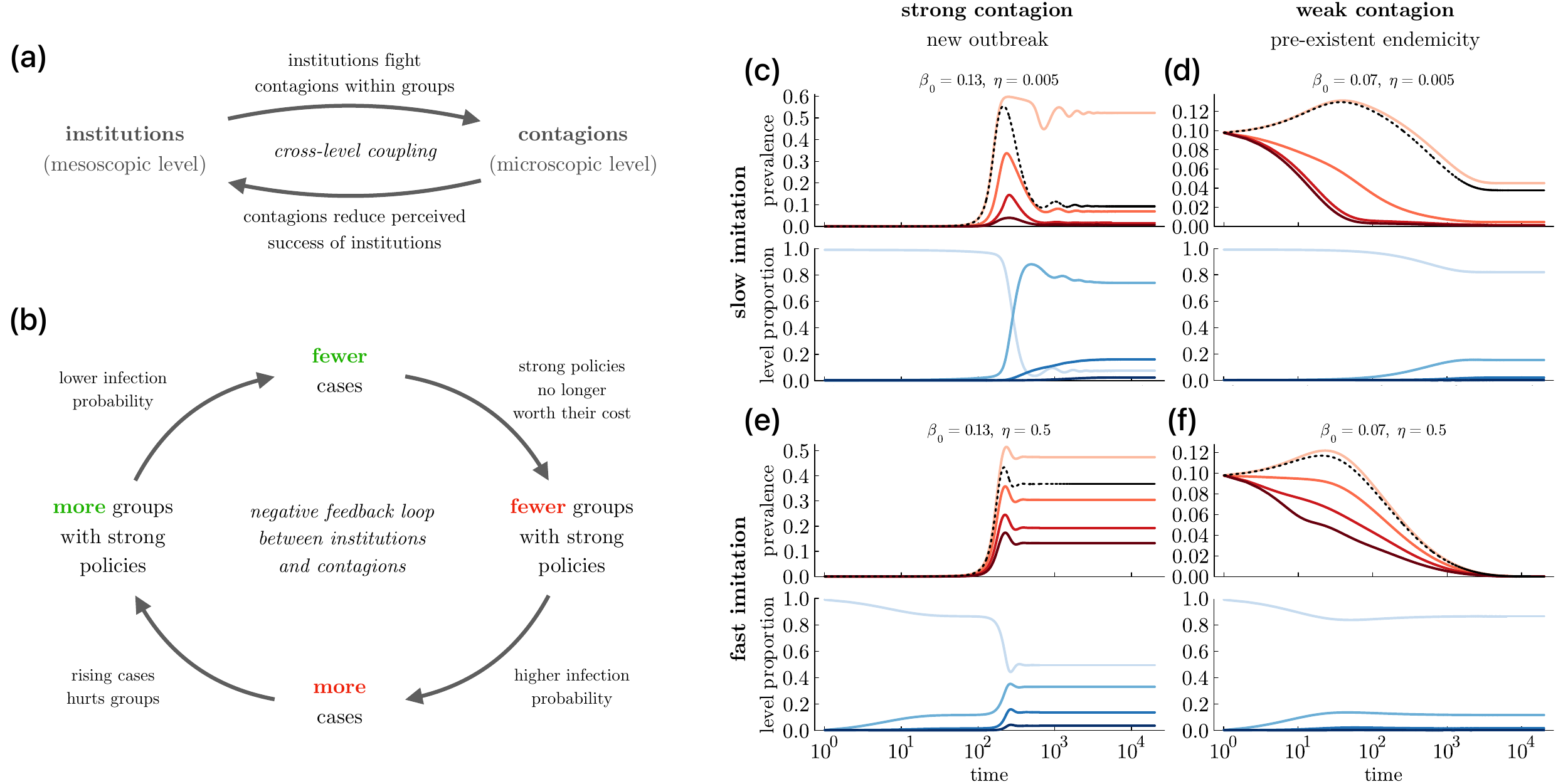}
    \caption{  \justifying
    \textbf{Cross-level mechanisms and multiple scenarios from the model.}  
    (a) Sketch of the cross-level coupling between institutions and contagions. (b) This coupling yields a negative feedback loop which generates the observed damped oscillatory dynamics. Considering either slow ($\eta=0.005$, (c, d)) or fast copying rate ($\eta=0.5$, (e, f)) and stronger ($\beta_0=0.13$, (c, e)) or weaker ($\beta_0=0.07$, (d, f)) contagions (together with different initial conditions), our model can reproduce a diversity of scenarios. Different shades of red indicate the prevalence within the different institutional levels, from $\ell=1$ (lightest shade) to $\ell=4$ (darkest shade), while the dotted line denotes the population average. Analogously, shades of blue indicate the proportion of each level. The system starts with $99\%$ of the groups at $\ell = 1$, the remaining portion being uniformly distributed among the other levels.
    (c, e) As for a new outbreak, only an infinitesimal fraction of the population is infected initially; $\beta_0 = 0.13$.
    (d, f) The initial prevalence is set equal to the equilibrium value it would attain for $\beta_0 = 0.07$ in absence of interventions (slightly less than $10\%$).
    The remaining parameters are fixed at $\rho = 0.05$, $\alpha=1$, $\gamma=1$, $b=-1$, $c=1$, and $\mu = 0.0001$. Notice that, since the system exhibits a single equilibrium state for each set of parameters, this state depends on $\eta$ and $\beta_0$, but not on the initial conditions.}
    \label{fig:temp_evo}
\end{figure*}

\subsection{Temporal analysis}
\label{sec:papertag.tempAnalysis}

By looking at the entire temporal evolution of the system, we see how the model is able to generate a variety of interesting scenarios. To observe this we vary either the values of the parameters, which determine the equilibrium distribution \{$G_{i,\ell}^\star\}$ of group states, and the initial conditions, which exclusively affect the transient behaviour. The existence of a single equilibrium state for the system is due to the stabilising negative feedback loop between institutional level and prevalence~\footnote{Multi-stability is instead induced by a positive feedback~\cite{hebert-dufresne_source-sink_2022}}: stronger institutions better avoid infections and so increase their fitness until infections are too few for such institutions to be worth their cost, leading their fitness to decrease again (see panels (a) and (b) in Fig.~\ref{fig:temp_evo} for a visualisation of the feedback).

We vary the value of the institutional copying rate, $\eta$, and of the basic infection rate, $\beta_0$, to explore the effect of the timescale separation between the (mesoscopic) institutional selection process and the (microscopic) contagion process. The coupled evolution of the two and their nonlinear nature prevent us from making a precise comparison of the two timescales. Nonetheless, we can safely qualify copying rate as either ``slow'' or ``fast'' compared to the contagion based on $\eta$ being either much smaller or larger than $\beta_0$, respectively. Any intermediate case can be understood as a combination of these two limit cases.

We initially seed either an infinitesimal fraction of the population, representing the case of a new (but not necessarily unknown) contagion, or a finite one, representing the case of an outbreak that reached endemicity without being (effectively) fought until then. In the latter case, the initial prevalence is set equal to the equilibrium value it would attain in absence of interventions (i.e., if only $\ell = 1$ existed). Moreover, we initialise the system with almost all the groups ($99\%$) in the lowest level, $\ell = 1$. In fact, whether the faced contagion has never been experienced or it is known but no measures have been deployed until then by a large majority of the groups, discovering, developing and implementing new interventions takes time. This leaves out the case of a new outbreak of a previously eradicated contagion, for which many groups may be ready to put in place strong interventions from the beginning. Nonetheless, as discussed below, such initial preparedness doesn't change the qualitative picture.

The results are reported in Fig.~\ref{fig:temp_evo}. A damped oscillatory transient is observed in all scenarios. The oscillations are indeed not an artifact of the model's details, but originate from the fundamental mechanism of mutual negative feedback between infection spread and institutional level the model accounts for. The ability of stronger policies to reduce infections increases their fitness, leading groups with weaker institutions to level up their policies. But these groups haven't yet had enough time to reduce infections among their members, consequently raising the prevalence associated to the upper institutional level they recently adopted and causing the fitness of the latter to lower. In turn, this leads some groups to level down their institutions and partially restart the cycle. Consistently, the higher the copying rate, $\eta$, the more frequent are the oscillations and the faster they wane, eventually becoming undetectable for very fast copying rate.

This temporal back and forth in the institutional space corresponds to variations of the infection rate within the groups, which also contribute to the oscillations observed in the level-specific and global prevalence. In particular, the first peak we observe is due to the time required for an unprepared population to explore stronger policies. Since the latter are worth their cost only when there are enough infections to be avoided, their implementation comes too late (more so when starting from an almost fully susceptible population, see Figs.~\ref{fig:temp_evo}~(c) and (e)), the exponential phase having already come to an end. Accordingly, the overshoot is experienced by groups at all institutional levels and is higher under slower copying rate (compare (c) to (e) and (d) to (f)). Notice that increasing the fraction of groups with ready-to-implement interventions at the start would solely result in a reduction of the amplitudes of the first oscillations (recall that the system has no dependence on the initial conditions).

\begin{figure}
  \centering	
    \includegraphics[width=\linewidth]{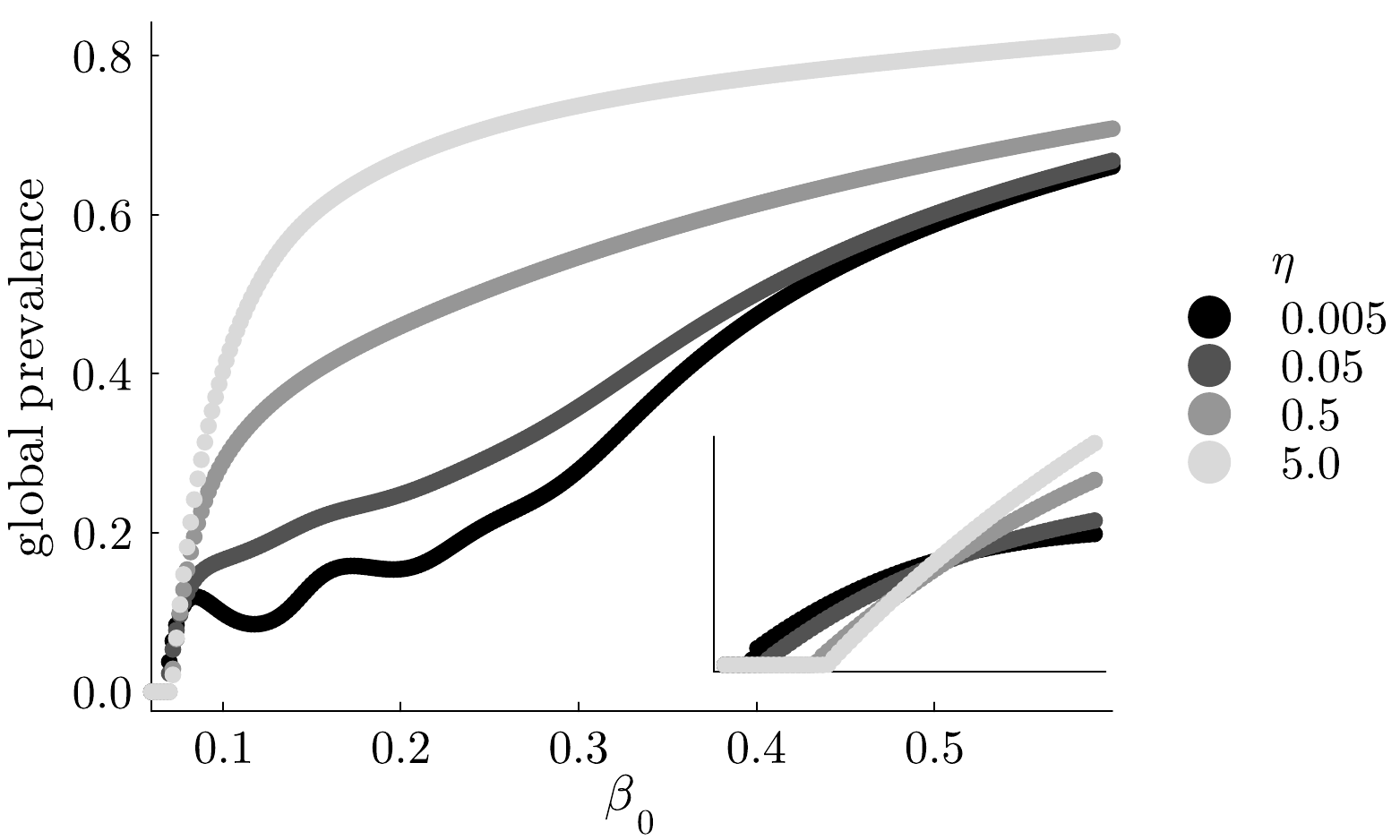}  
  \caption{ \justifying
  \textbf{Effect of institutional copying rate on equilibrium global prevalence.}
  Phase diagram representing the equilibrium global prevalence against the basic infection rate, $\beta_0$, for different values of the institutional copying rate, $\eta$. The inset plot zooms in the critical region. The remaining parameters are fixed at $\rho = 0.05$, $\alpha=1$, $\gamma=1$, $b=-1$, $c=1$, and $\mu = 0.0001$. A faster copying rate yields (much) larger outbreaks, while slightly increasing the epidemic threshold. Under slow enough copying rate, a non-monotonic behaviour is observed.
  }
  \label{fig:call4action}
\end{figure}

The different combinations of contagiousness and copying rate speed generate the four scenarios displayed in Fig.~\ref{fig:temp_evo}. In the presence of a strong contagion (panels (c) and (e)), i.e., one which is too infectious to be eradicated, only a slow copying rate is able to greatly reduce the global prevalence. Since in this case the groups stay long with their current policies before considering others, stronger institutions have enough time to reduce infections and prove they are worth their cost. Higher institutional levels are thus selected, leading, after a large first wave, to a (very) low endemicity either within those levels and globally. We observe this in panel (c), where the largely most chosen level is $\ell = 2$ eventually, followed by level $\ell = 3$, whereas only a small proportion of groups does not implement any intervention ($\ell = 1$), suffering in this way a high endemicity. As shown in panel (e), a fast copying rate instead impedes the selection of stronger policies and the system rapidly settles in a state of high prevalence at all levels.

Against a weak contagion (panels (d) and (f)), i.e., one which is slightly above the epidemic threshold in the absence of interventions, the benefit of waiting is lost. Since endemicity is low even with no interventions, the lower prevalence attained by groups implementing some control policies does not justify the cost of the latter. Whatever the copying rate speed, groups therefore opt to not cover any cost for most of the time and simply let the contagion spread. Surprisingly, only a fast enough copying rate is able to lead to eradication in this case and, differently from before, the key is not favouring stronger policies---which, indeed, are always rarer the faster is copying rate. A fast selection allows groups to frequently explore other levels and avoid to stagnate for long times in the state of no interventions ($\ell = 1$) where the contagion is free to grow exponentially and revive, as indeed occurs under slow copying rate (panel (d)). Thanks to the frequent switching, after a short transient, the outbreak becomes overall subcritical and dies out eventually (panel (f)).

\subsection{Equilibrium analysis}
\label{sec:eq}

We focus now on the equilibrium properties of the model by looking at the values reached by the global and level-specific prevalence and by the levels' proportions.

The analysis of the temporal evolution already hinted at how the equilibrium state is closely and non-trivially tied to the institutional copying rate, $\eta$. In Fig.~\ref{fig:call4action}, we show how varying $\eta$ impacts the global prevalence as we increase the basic infection rate, $\beta_0$. A slower copying rate (smaller $\eta$) is beneficial to control the contagion, for stronger policies are given enough time to effectively reduce infections and prove to be worth their cost, being eventually selected by most of the groups. This holds except near eradication, where instead a fast enough copying rate leads to eradication but a slower one doesn't (see the inset plot), as discussed before. In summary, a higher copying rate causes a (much) higher endemicity, while barely increasing the epidemic threshold.

\paragraph*{Call for action.}

\begin{figure}
  \centering	
    \includegraphics[width=0.98\linewidth]{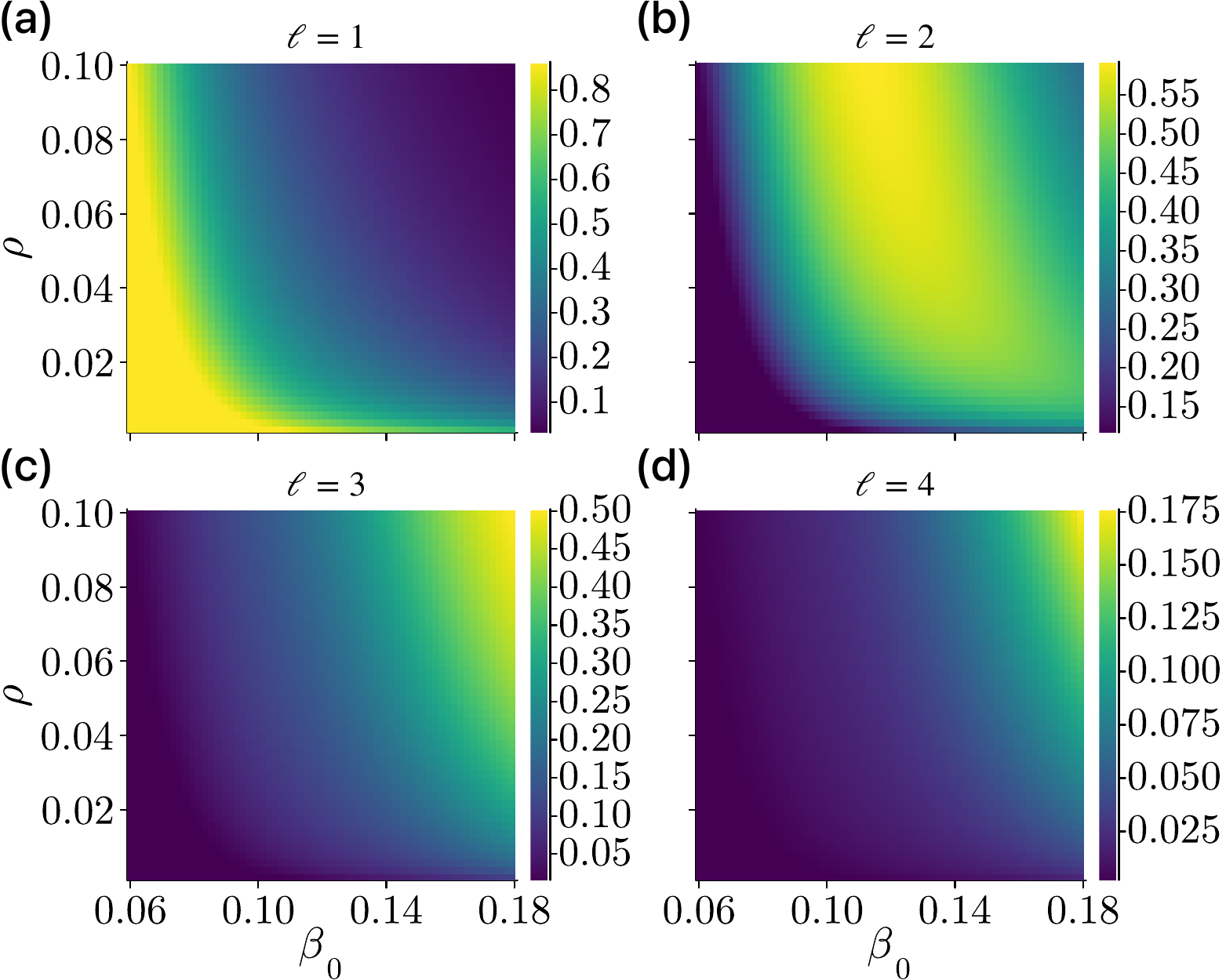}  
  \caption{ \justifying
  \textbf{Institutional localisation.} Heatmaps representing the proportion of groups with institutional level $\ell\in\{1,2,3,4\}$ (from (a) to (d), respectively) at equilibrium as a function of the basic infection rate $\beta_0$ and the inter-group coupling $\rho$.
  The remaining parameters are fixed at $\eta = 0.05$, $\alpha=1$, $\gamma=1$, $b=-1$, $c=1$, and $\mu = 0.0001$. The system localises at increasingly higher institutional levels in response to increased infectivity and/or inter-group coupling.
  }
  \label{fig:localisation}
\end{figure}

Interestingly, slowing down copying rate enough, we observe the prevalence curve becoming non-monotonic, with multiple local minima. There can be at most $L-1$ of them, as many as the number of levels implementing some interventions ($\ell = 2, \dots, L$). The minima are due to an \textit{institutional localisation}, according to which different regions of the parameter space are dominated by specific institutional levels containing the majority of groups~\cite{hebert-dufresne_source-sink_2022}, as shown in Fig.~\ref{fig:localisation}. Such levels are the fittest in that region.  Weakest institutions ($\ell = 1$) dominate the fitness landscape in regimes of low $\beta_0$ and (very) low $\rho$, for stronger policies are not selected in the presence of few infections. As we increase $\beta_0$ (or $\rho$, when enough above the epidemic threshold), however, stronger institutions prove their ability to control the contagion, being increasingly selected.

\begin{figure}
  \centering	
    \includegraphics[width=\linewidth]{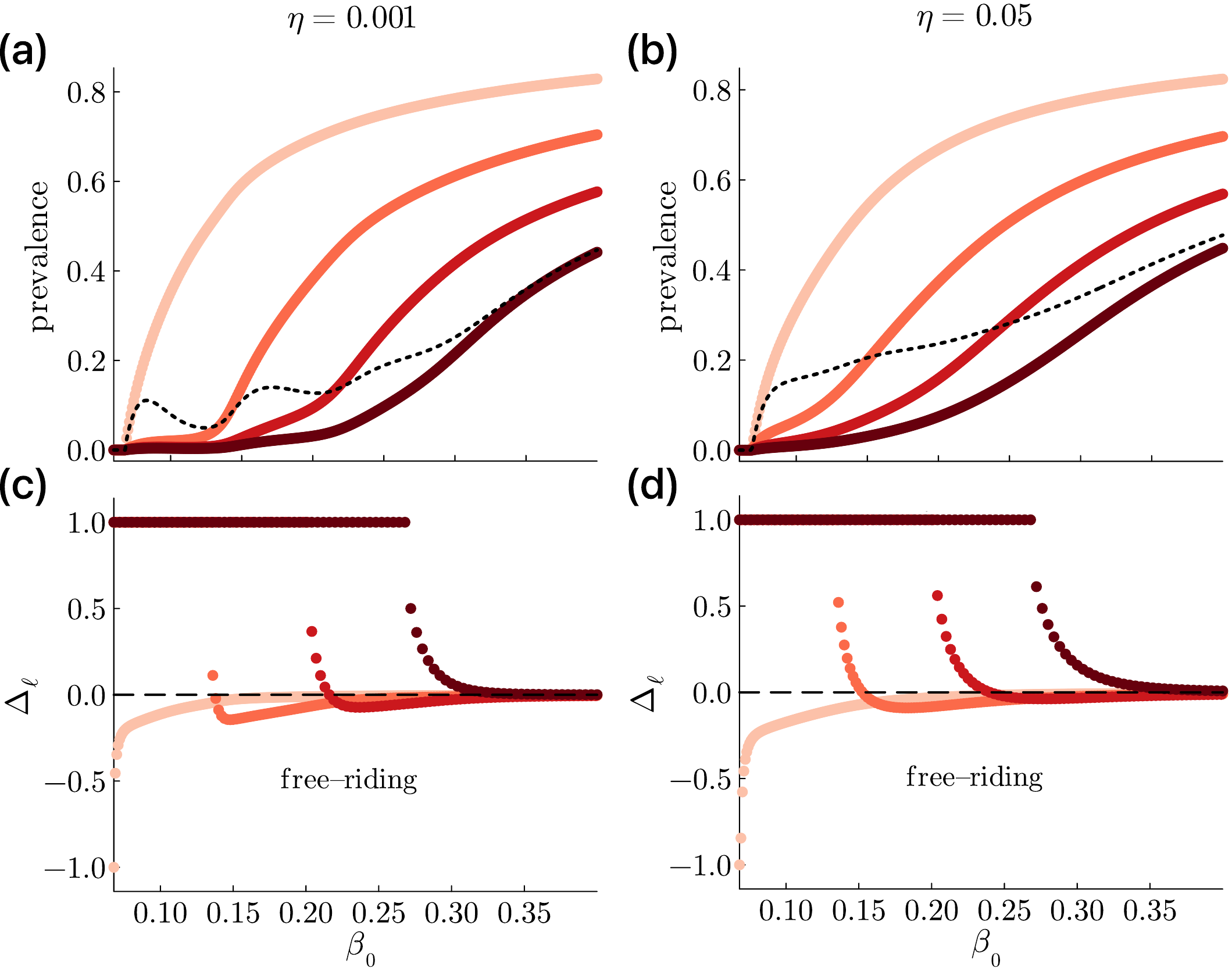}  
  \caption{ \justifying
  \textbf{Equilibrium prevalence and free-riding onset.}
  (a)-(b) Level-specific equilibrium prevalence $I_\ell$, $\ell\in\{1,2,3,4\}$ (from the lightest to the darkest shade, respectively), and global prevalence (dotted line) versus the basic infection rate $\beta_0$, for copying rate $\eta=0.001$ and $0.05$; $\rho = 0.05$. (c)-(d) Normalised difference $\Delta_\ell = (I_\ell - \tilde{I}_\ell)/(I_\ell + \tilde{I}_\ell)$ between $I_\ell$ and $\tilde{I}_\ell$, the latter being the equilibrium prevalence obtained if only institutional level $\ell$ was allowed. A negative (positive) value of $\Delta_\ell$ indicates that groups of institutional level $\ell$ are experiencing a lower (higher) prevalence by free-riding (suffering) the (lack of) efforts from groups of higher (lower) level. The remaining parameters are fixed at $\alpha=1$, $\gamma=1$, $b=-1$, $c=1$, and $\mu = 0.0001$.
  }
  \label{fig:free_riding}
\end{figure}

In fact, the localisation occurs for any value of the copying rate, yet only under slow copying rate it becomes strong enough to produce a non-monotonic dependence of the prevalence on $\beta_0$ (see the interactive Supplementary Material to explore how localisation is modulated by varying the copying rate~\footnote{Interactive Supplementary Material: \href{https://observablehq.com/@joint-lab/call-for-action}{https://observablehq.com/@joint-lab/call-for-action}}). This is explained by looking at how the different levels behave. In Figs.~\ref{fig:free_riding}~(a) and (b) we show how the level-specific equilibrium prevalence, $I_\ell = \sum _i (i/n) G_{i,\ell} / \sum _i G_{i,\ell}$, changes with $\beta_0$ for different values of $\eta$. As one may expect, the higher the institutional level $\ell$, the higher is the value of $\beta_0$ at which $I_\ell$ departs significantly from zero. In particular, at the typical time at which selection occurs (1/$\eta$), groups implementing some interventions ($\ell\in\{2,3,4\}$) will have on average less infected members than those groups implementing none ($\ell = 1$), and more so the stronger the interventions are. This prevalence gap will be also wider the more infectious is the contagion (notice that $\partial\beta_\ell/\partial\beta_0 = \ell^{-\alpha}$ decreases with $\ell$) and the slower is copying rate, for groups spend more time in their current institutional level. If a higher level is relatively not too costly, the resulting gap in fitness will then favour the selection of this level, producing a major localisation. Higher levels are eventually preferred, leading to lower endemicity either within those levels and globally (panel (a)). We recognise this phenomenon as a \emph{call for action}, as groups are called upon to invest in more effective policies to control a stronger contagion. To note that the remaining lowest-level groups are instead worse off under slow copying rate, for they spend more time without interventions, eventually growing more infections. 

In the end, the existence of such non-monotonic regime, in which three or more values of the basic infection rate, $\beta_0$, correspond to the same equilibrium prevalence, fully display the role of the institutional dimension in the co-evolutionary setup. Knowledge of this dimension is generally needed to correctly read the system.

\paragraph*{Institutional free-riding.}

Similar to individuals, groups with weaker institutions can (not necessarily voluntarily) take advantage of groups with stronger institutions by enjoying a lower prevalence without contributing to the efforts or costs required for implementing stronger policies. To quantify this, we introduce $\Delta_\ell = (I_\ell - \tilde{I}_\ell) / (I_\ell + \tilde{I}_\ell)$, a level-specific measure taking values in $[-1,1]$. This is the normalised difference between $I_\ell$, obtained when all the $L$ institutional levels are accessible, and $\tilde{I}_\ell$, the equilibrium prevalence when only level $\ell$ exists.
We say that groups of level $\ell$ free-ride when $\Delta_\ell<0$, i.e., when $I_\ell < \tilde{I}_\ell$, while $\Delta_\ell>0$ indicates they are worse off, being $I_\ell > \tilde{I}_\ell$. The closer $\Delta_\ell$ is to $-1$ (resp., $1$) the more the groups at level $\ell$ are enjoying (suffering) the (lack of) efforts from groups of higher (lower) level~\footnote{Notice that since $\Delta_\ell$ is a measure of relative change within level $\ell$, $\Delta_\ell < \Delta_{\ell^\prime} < 0$ ($\Delta_\ell > \Delta_{\ell^\prime} > 0$), $\ell \neq \ell^\prime$, doesn't necessarily mean that groups of level $\ell$ see their prevalence lowered (raised) in absolute terms more than those of level $\ell^\prime$ do. It is rather more informative to look for when $\Delta_\ell$ and $\Delta_{\ell^\prime}$ have opposite signs}.

In Figs.~\ref{fig:free_riding}~(c) and (d), we show how $\Delta_\ell$ changes with $\beta_0$. Differently from the equilibrium prevalence $I_\ell$, which behaviour crucially depends on the rate of institutional copying rate, $\Delta_\ell$ displays a qualitatively general behaviour which can be summarised as follows. Groups in the lowest institutional level, $\ell = 1$, are always better off in the presence of groups in higher levels, i.e., they free-ride ($\Delta_1 < 0$). On the contrary, groups in the highest level, $\ell = 4$, are always worse off ($\Delta_4 > 0$) due to the interaction with groups at lower levels. Groups with intermediate institutions ($\ell \in \{2, 3\}$) suffer the presence of lower-level groups, but also enjoy the presence of higher-level ones. Then, as long as the infection rate $\beta_0$ is enough smaller than the epidemic threshold for level $\ell \in \{2, 3\}$ (i.e., the value at which $\tilde{I}_\ell$ becomes positive), groups at this level experience some endemicity which they would not if lower levels were absent, implying $\Delta_\ell = 1$. However, increasing $\beta_0$ above the respective epidemic threshold, $\Delta_\ell$ decreases, up to the point at which it crosses zero, marking the onset of the free-riding regime for that level ($\Delta_\ell < 0$). Lastly, since $I_\ell$ and $\tilde{I}_\ell$ must converge at very high infectivity, $\Delta_\ell$ approaches zero from below for $\ell \in \{1, 2, 3\}$ and from above for $\ell = 4$. A completely analogous analysis holds when varying $\rho$ in terms of its critical value for each given $\ell$.

\section{Discussion}
\label{sec:papertag.concludingremarks}

\paragraph*{Reproducing real-world scenarios.}

To summarize our model dynamics, we can establish a qualitative correspondence between the scenarios produced by the model, as illustrated in Fig.~\ref{fig:temp_evo}, and some real-world contagions. We do not attempt to fit our model or observe local group policies and the relevant copying rates. We simply focus on the general dynamics of global incidence of different contagions to draw analogies between the model and known epidemics or harmful behaviours. 

Under slow copying rate, stronger institutions are selected through time. 
After an initial wave of cases, the contagion is partially controlled and settles to a low global endemicity. Groups adopting strong institutions achieve lower endemicity, while this remains high where interventions are absent (panel (c)). 
This mirrors what we understand qualitatively from cigarette consumption, where there was a significant increase in cigarette sales up to the 1950s, followed by a decline due to institutional policies such as federal taxes and in-flight smoking bans \cite{wong2023covid, caplow_first_2000}. Further slowing down copying rate, we find that increasingly stronger institutions are preferred. The first wave becomes larger but the global equilibrium prevalence is eventually lower, potentially approaching eradication. 
This resembles the cases of HIV and Syphilis~\cite{caplow_first_2000}.

Under fast copying rate, cheap but weak institutions are selected, letting an outbreak to become highly endemic. Even the strongest institutions are unable to achieve low endemicity (panel (e)), as for chlamydia~\cite{centers_for_disease_control_and_prevention_sexually_2014}.

Only when the copying rate is not too slow, slight interventions suffice to control and eventually globally eradicate a weakly infectious contagion (which would become lowly endemic in the absence of interventions; (panel (f)), as for diphtheria and pertussis~\cite{caplow_first_2000}; otherwise, the contagion persists (panel (d)), as for marijuana, cocaine, and hallucinogen usage~\cite{caplow_first_2000}. We note that substances like marijuana introduce an additional layer in the co-evolution of policies and behaviours, as institutions possess the capacity to influence both behaviour and public perceptions surrounding it. Even though marijuana may not have the same health impacts as diseases or tobacco, institutions can choose to make the drug illegal, thus increasing the penalties associated with its use. Investigating how institutions can change the rules of the game by modifying the payoffs of behaviours represents a future research direction within our group-based formalism.

\paragraph*{The role of group heterogeneity.}

We showed how the overall contagion dynamics is strongly affected by the rate at which groups adapt their policies. This begs the question of what would be the effect of having heterogeneous copying rates across groups. 
Additionally, group heterogeneity of various kinds may affect the system; one might envision that some groups experience disproportionate penalties of having infected members, such as states or countries with more elderly population~\cite{wong2023covid}. Or policy mandates may entail heterogeneous costs based on the infrastructure priors to the contagion (for example, implementing a stronger vaccination policy is cheaper for countries with pharmaceutical infrastructure~\cite{Working_Group_2021}). 

\paragraph*{Reintroducing a global top-down influence.}
One of the inspirations for constructing our model as it is was the absence of bottom-up governance models. 
However, institutional theorists might add that another realistic step is the centralised coordination of groups from a higher-level institution~\cite{ostrom_governing_1990}, such as a government asking local organisations to impose a mask mandate.
In our model this could be simply implemented via a transition rate towards stronger policies which is function of the number of (newly) infected individuals overall. 

\paragraph*{Conclusion.}

Our model illustrates that solely focusing on individual behaviours, without taking into account the response of institutions and other groups, would miss important group-level feedback mechanisms. First, one group improving its own situation through institutional policies can inspire other groups to do the same. Second, higher transmission rates can increase the need for institutions more than it increases the contagion, resulting in lower prevalence levels. To fully understand the efficacy of policies, it is therefore crucial to examine the diffusion and feedback of policies across different groups. 

The bottom line is that group dynamics and institutions play an important part in the dynamics of epidemics, misinformation, pests and other contagions. Yet most models of contagion ignore cultural group selection and other mechanisms of group-level dynamics. We thus hope that future work will further develop co-evolutionary group-based models by accounting for additional organisational scales or heterogeneities across and within groups.

\acknowledgments

This work was partially supported by the Sloan Foundation through the Vermont Research Open Source Program Office (J.S.-O. \& L.H.-D.), by the European Union's Horizon 2020 research and innovation program under the Marie Sk\l{}odowska-Curie Grant Agreement No.\ 945413 (G.B.), by the National Science Foundation award EPS-2019470 (T.W. \& L.H.-D.), by the National Institute of Food and Agriculture project no. VT-0095CG (S.F.R. \& L.H.-D.), and by the National Institutes of Health 1P20 GM125498-01 Centers of Biomedical Research Excellence Award (L.H.-D.)

\end{document}